\documentclass[journal]{IEEEtran}
\usepackage{amsmath}
\usepackage{subfigure}
\usepackage{graphicx}
\usepackage{latexsym}
\usepackage{amssymb,amsbsy,verbatim,array}
\usepackage{epstopdf}
\usepackage{cite}
\usepackage{color} 
\usepackage[colorlinks,citecolor=blue,linkcolor=blue,urlcolor=blue]{hyperref}
\newtheorem{remark}{Remark}

\def\rb{\rho_d}
\def\ra{\rho_0}

\def\ga{\mathbf{Ga}}
\def\dj{\delta_j}

\everymath{\displaystyle}\begin{document}
\title{FAS-RIS for V2X: Unlocking Realistic Performance Analysis with Finite Elements}

\author{Tuo Wu, Xiazhi Lai, Kangda Zhi, Maged Elkashlan, Naofal Al-Dhahir, \emph{Fellow, IEEE},\\ Matthew  C. Valenti, \emph{Fellow}, \emph{IEEE}, and Fumiyuki Adachi, \IEEEmembership{Life Fellow,~IEEE}
	\thanks{(\textit{Corresponding author:Xiazhi Lai.})}
\thanks{T. Wu is with the School of Electronic and Information Engineering, South China University of Technology, Guangzhou 510640, China (E-mail: $\rm  wtopp0415@163.com$).   X. Lai is with the School of Computer Science, Guangdong University of Education, Guangzhou, Guangdong, China (E-mail: $\rm xzlai@outlook.com$).  K. Zhi is with Communications and Information Theory Group (CommIT), Technische Universitat Berlin, 10587 Berlin, Germany (E-mail: $\rm k.zhi@tu$-$\rm berlin.de$). M. Elkashlan is with the School of Electronic Engineering and Computer Science at Queen Mary University of London, London E1 4NS, U.K. (E-mail: $\rm maged.elkashlan@qmul.ac.uk$). Naofal Al-Dhahir is with the Department of Electrical and Computer Engineering, The University of Texas at Dallas, Richardson, TX 75080 USA (E-mail: $\rm aldhahir@utdallas.edu$).    M. C. Valenti is with the Lane Department of Computer Science and Electrical Engineering, West Virginia University, Morgantown, USA (E-mail: $\rm valenti@ieee.org$). F. Adachi is with the International Research Institute of Disaster Science (IRIDeS), Tohoku University, Sendai, Japan (E-mail: $\rm adachi@ecei.tohoku.ac.jp$).} 
}
\markboth{}
{Lai \MakeLowercase{\textit{et al.}}: }

\maketitle

\thispagestyle{empty}

\begin{abstract}
The synergy of {fluid antenna systems (FAS)} and {reconfigurable intelligent surfaces (RIS)} promises robust {vehicle-to-everything (V2X)} links, yet most analyses invoke the {central limit theorem (CLT)} and thus fail to capture practical, finite-size deployments. This paper develops a realistic and tractable framework for FAS--RIS systems with a finite number of elements. We approximate the cascaded end-to-end gain via a Gamma distribution using moment matching, model spatial dependence across FAS ports with a block-correlation structure, and derive an accurate, closed-form approximation for the outage probability using Gauss--Chebyshev quadrature. Extensive simulations show that the proposed Gamma-based analysis markedly outperforms CLT-based baselines—especially for small numbers of RIS elements and ports—while converging to them as the array grows. The results provide actionable guidance for V2X design and dimensioning under practical constraints on RIS size, FAS aperture, and channel correlation.
\end{abstract}

\section{Introduction}
To meet the increasing demand for high data rates in vehicular networks, particularly in {vehicle-to-everything (V2X)} communications that enable vehicles to exchange information with surrounding infrastructure, pedestrians, and other vehicles, the {fluid antenna system (FAS)} has emerged as one of the crucial techniques for the next generation of wireless communications. Due to the high mobility of vehicles and the complex propagation environment inherent in V2X scenarios, conventional fixed multiple antenna techniques face limitations in providing sufficient spatial diversity, especially within the compact size of vehicle-mounted devices. FAS, however, can flexibly switch its antenna port to the most favorable location, thereby significantly enhancing signal reception quality \cite{FAS20,Wong-frontiers22,MFAS23,TWu25}. In practice, the FAS can be implemented via pixel-based or liquid metal structures \cite{Rodrigo14,Huang21}, and the optimal port can be selected via learning-based methods \cite{Chai22,Waqar23,Yang25}. FAS has also shown potential in integrated sensing and communications (ISAC) and unsourced random access scenarios \cite{Zhang25JSAC,Zhang25WCL}. To evaluate the performance of FAS-enabled communications, a simplified channel correlation model has been proposed in \cite{FAS21,FAS22}, where the aggregated effect of port correlation was considered, and the outage performance of multiple access networks has been analyzed \cite{FAMS,FAMS23}. A more accurate Jakes' correlation model has been analyzed in \cite{Khammassi23}, while the difficulty of multiple-dimensional integral makes the analysis intractable for other complicated scenarios. Hence, a block-correlation approximate model has been proposed to maintain accuracy and keep the analysis tractable \cite{BC24,Lai25JSAC,WuTuo25,Zheng25}.

In urban vehicular networks, the direct link between the base station and a vehicle is often blocked by buildings and other obstacles, leading to deep fading. To cope with this challenge, the {reconfigurable intelligent surface (RIS)} technique has been proposed. By intelligently reshaping the phase of radio-frequency (RF) signals, an RIS can establish a reliable alternative link and bring extra diversity to improve the quality of received signals for the vehicle. To this end, the coverage probability of RIS networks has been evaluated in \cite{Yang20}, and the outage probability of the RIS network has been provided as well.
Moreover,  \cite{Gan21} considered   RIS networks in the presence of multiple users, and quantified their performance terms of ergodic capacity. RIS has also been explored in cell-free massive MIMO systems \cite{ShiProc,Shi25TWC}.

To further enhance the performance of V2X communications, the synergy of FAS and RIS has been explored in several pioneering works \cite{LaiX242,YaoJ251}, while RIS has also been applied to localization and security scenarios \cite{Wu25RIS}. However, these foundational analyses often rely on the {central limit theorem (CLT)} to model the channel, an approach that is only accurate for a large number of RIS reflecting elements. This idealized assumption creates a critical gap between theoretical predictions and the performance of realistic V2X systems, where physical and cost constraints often dictate the use of small-scale RIS. For instance, to ensure cost-effective, wide-scale coverage in V2X deployments, RISs are likely to be deployed on existing urban infrastructure like lampposts or building facades, where the number of possible elements is inherently finite. In these practical scenarios, CLT-based approximations can lead to significant errors, hindering effective system design. The primary goal of this paper is to bridge this gap, unlocking a more realistic and accurate framework for performance analysis. To this end, this paper makes the following key contributions:
\begin{itemize}
	\item \textbf{Gamma-based Analytical Framework:} We propose a novel analytical framework based on the Gamma distribution to characterize the end-to-end channel of a downlink FAS-RIS aided vehicular network. This approach overcomes the limitations of the conventional CLT-based method, especially in practical scenarios with a limited number of RIS elements.
	\item \textbf{Closed-form Outage Probability Expression:} We derive a new closed-form approximate expression for the outage probability of the considered system. The derivation leverages the Gamma distribution approximation in conjunction with a block-correlation channel model, providing a tractable way to evaluate system performance.
	\item \textbf{Performance Validation and Insights:} We provide extensive numerical results to validate the accuracy of our proposed analysis. The results confirm that our Gamma-based approach significantly outperforms the CLT-based benchmark, particularly when the number of RIS elements is small (e.g., fewer than 30), offering valuable insights for the practical design of FAS-RIS systems in vehicular networks.
\end{itemize}

\emph{Notation}: $X\sim\mathcal{CN}(\alpha,\beta)$ denotes a complex Gaussian random variable (RV) with mean $\alpha$ and variance $\beta$. $\mathbf{E}(\cdot)$ and $\mathbf{Var}(\cdot)$  denote the expectation and variance of an random variable (RV), respectively. $f_X(x)$ and $F_X(x)$ denote the probability density function (PDF) and cumulative distribution function (CDF), respectively.
\begin{figure}[t]
	\centering
	\includegraphics[width=0.9\linewidth]{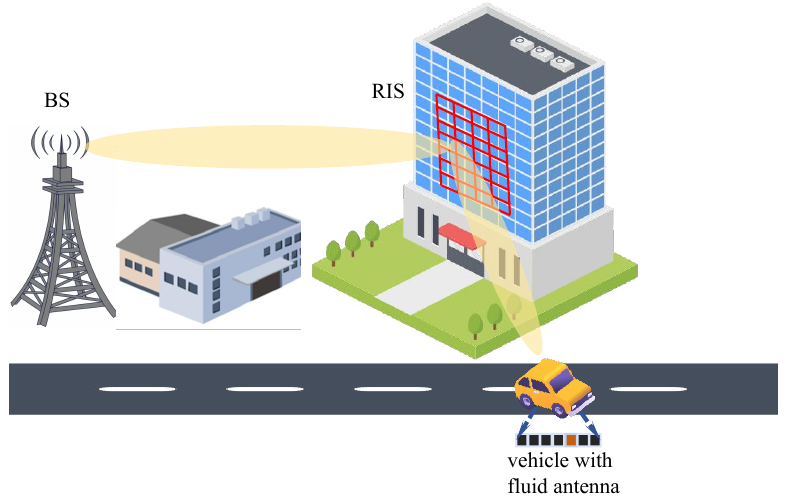}\\
	\caption{The system model of  FAS-RIS aided vehicular network.}\label{fig1}
\end{figure}

\section{System Model} 

We consider an FAS-RIS-aided vehicular downlink communication network, as depicted in Fig. \ref{fig1}.\footnote{{The current model neglects Doppler-induced correlation drift due to high-mobility effects; future work will integrate mobility-aware channel evolution to capture highway-speed V2X deployments.}} A single-antenna base station (BS) serves a vehicle equipped with a linear FAS. The direct link between the BS and the vehicle is assumed to be blocked by obstacles, e.g., buildings. To establish a reliable communication link, an RIS composed of $M$ reflecting elements is deployed to assist the downlink transmission. The FAS at the vehicle consists of $K$ available ports distributed over a line of length $W\lambda$, where $W$ is a normalization factor and $\lambda$ is the carrier wavelength. {Here, $W$ denotes the system bandwidth used for normalizing the spectral efficiency; all rates expressed in bit/s/Hz are referenced to this bandwidth.} The port that offers the maximum channel gain is selected for data reception.

\subsection{Channel and Signal Model}
The end-to-end communication from the BS to the $k$-th port of the FAS comprises two hops: the BS-RIS link and the RIS-vehicle link.
\subsubsection{Channel Model}
The channel coefficient of the link between the BS and the $m$-th RIS element is denoted by $h_m \sim \mathcal{CN}(0, \epsilon_1)$. The channel coefficient between the $m$-th RIS element and the $k$-th FAS port is denoted by $v_{m,k} \sim \mathcal{CN}(0, \epsilon_2)$. Consequently, the channel gains $|h_m|$ and $|v_{m,k}|$ follow a Rayleigh distribution. Due to the continuous nature of the FAS, the channels to its different ports are spatially correlated. We adopt the widely-used Jakes' model to characterize this correlation. The correlation coefficient between the channels to ports $k$ and $l$, for any RIS element $m$, is given by
\begin{align}\label{a4}
g_{k,l}=J_0\bigg(\frac{2\pi |k-l| W}{K-1}\bigg),
\end{align}
where $J_0(\cdot)$ is the zeroth-order Bessel function of the first kind. The channel vector from the $m$-th RIS element to all $K$ FAS ports, $\mathbf{v}_m = [v_{m,1}, \dots, v_{m,K}]^T$, is a complex Gaussian vector with zero mean and a covariance matrix $\mathbf{\Sigma} \in \mathbb{C}^{K \times K}$. Given the distance-dependent nature of Jakes' model, $\mathbf{\Sigma}$ is a Toeplitz matrix with $[\mathbf{\Sigma}]_{k,l} = g_{k,l}$.

\subsubsection{Signal Model}
Let $s$ be the transmitted symbol from the BS with average power $\mathbf{E}[|s|^2]=1$, and let $P_S$ be the transmit power. The signal received at the $k$-th FAS port is the superposition of the signals reflected from all $M$ RIS elements, which can be expressed as
\begin{align}\label{a1}
y_{k}=\sqrt{P_S} \left( \sum_{m=1}^{M} h_m v_{m,k} e^{i\theta_m} \right) s + n_k,
\end{align}
where $n_k \sim \mathcal{CN}(0, \sigma_n^2)$ is the additive white Gaussian noise (AWGN) at the receiver, and $\theta_m$ is the phase shift applied by the $m$-th RIS element. To maximize the received signal strength, the RIS is configured to perform optimal co-phasing by aligning the phases of all cascaded paths. This is achieved by setting the phase shift to $\theta_m = -\arg(h_m v_{m,k})$. {Selecting $\theta_m = -\arg(h_m)$ aligns the phases of the scattered components, converting the complex-valued sum into a coherent accumulation of magnitudes and thus maximizing the post-RIS gain.} With this optimal configuration, the effective channel gain for the $k$-th port becomes the coherent sum of the individual path gains:
\begin{align}\label{a2}
\psi_k = \sum_{m=1}^{M} |h_m| |v_{m,k}|.
\end{align}
The FAS then selects the port with the maximum channel gain to decode the signal. The resulting channel gain is
\begin{align}\label{a3}
\psi^* = \max_{k \in \{1, \dots, K\}} \psi_{k}.
\end{align}

\subsection{Channel Statistics and Correlation}
To analyze the system's outage performance, we first characterize the statistical properties of the effective channel gain $\psi_k$.
\subsubsection{Mean and Variance}
Since $|h_m|$ and $|v_{m,k}|$ are independent Rayleigh-distributed random variables (RVs), the expectation of their product is $\mathbf{E}[|h_m||v_{m,k}|] = \mathbf{E}[|h_m|] \mathbf{E}[|v_{m,k}|] = (\pi/4)\sqrt{\epsilon_1\epsilon_2}$. The variance can be computed as $\mathbf{Var}(|h_m||v_{m,k}|) = \mathbf{E}[|h_m|^2|v_{m,k}|^2] - (\mathbf{E}[|h_m||v_{m,k}|])^2 = \epsilon_1\epsilon_2(1-\pi^2/16)$.
As $\psi_k$ is the sum of $M$ independent and identically distributed (i.i.d.) RVs $|h_m||v_{m,k}|$, its mean $E_\psi$ and variance $V_\psi$ are given by
\begin{align}\label{c1}
E_{\psi} &= M\mathbf{E}[|h_m| |v_{m,k}|] = \frac{M\pi\sqrt{\epsilon_1\epsilon_2}}{4},\\ \label{c2}
V_{\psi} &= M\mathbf{Var}(|h_m| |v_{m,k}|) = M\epsilon_1\epsilon_2\left(1-\frac{\pi^2}{16}\right).
\end{align}
\subsubsection{Correlation Coefficient}
The effective channels $\psi_k$ and $\psi_l$ for different ports $k$ and $l$ are correlated due to the spatial correlation of the RIS-vehicle channels. The Pearson correlation coefficient between them, denoted by $\eta(g_{k,l})$, is defined as
\begin{align}\label{c3}
\eta(g_{k,l}) &= \frac{\mathbf{E}[\psi_k\psi_l]-E_{\psi}^2}{V_{\psi}}\notag\\
&=\frac{M\epsilon_1\mathbf{E}[|v_{m,k}||v_{m,l}|]-E_{\psi}^2/M}{V_{\psi}},
\end{align}
where the derivation relies on the independence of channels across different RIS elements. The term $\mathbf{E}[|v_{m,k}||v_{m,l}|]$ is the expected value of the product of two correlated Rayleigh RVs, given by
\begin{align}\label{c4}
\mathbf{E}[|v_{m,k}||v_{m,l}|] = \int_{0}^{\infty}\int_{0}^{\infty}xyf_{|v_{m,k}|,|v_{m,l}|}(x,y)dxdy.
\end{align}
Here, $f_{|v_{m,k}|,|v_{m,l}|}(x,y)$ is their joint probability density function (PDF), which is given by
\begin{align}\label{c5}
f_{|v_{m,k}|,|v_{m,l}|}(x,y)=&\frac{4xy}{\epsilon_2^2(1-|g_{k,l}|^2)}e^{-\frac{x^2+y^2}{\epsilon_2(1-|g_{k,l}|^2)}}\notag\\
&\times
I_0\bigg(\frac{2|g_{k,l}|xy}{\epsilon_2(1-|g_{k,l}|^2)}\bigg),
\end{align}
where $I_0(\cdot)$ is the modified Bessel function of the first kind. { \eqref{c4} integrates the element-wise correlation kernel over the joint distribution of the Rayleigh magnitudes; evaluating this integral with the PDF in \eqref{c5} yields the closed-form expectation $\mathbf{E}[|v_{m,k}||v_{m,l}|]$, which is then substituted into \eqref{c3} to obtain the final correlation coefficient $\eta(g_{k,l})$.} This integral is typically computed via numerical methods.
Based on the Pearson correlation coefficient,  the correlation coefficient matrix of $\mathbf{\Psi}=[\psi_1,\cdots,\psi_K]$ is given as
\begin{align}\label{a6}
\mathbf{\Omega}\in \mathbb{R}^{K\times K} & =\mathbf{topelitz}\big(\eta(g_{1,1}),\eta(g_{1,2}),\cdots, \eta(g_{1,K})\big).
\end{align}
{The correlation matrix $\mathbf{\Omega}$ is Toeplitz, constructed from the spatial correlation vector $[\eta(g_{1,1}), \eta(g_{1,2}), \ldots, \eta(g_{1,K})]$, so that $[\mathbf{\Omega}]_{i,j} = \eta(g_{|i-j|+1,1})$ for $i,j \in \{1,\ldots,K\}$.}

\section{Outage Performance Analysis}
The outage probability is a key metric for evaluating the reliability of the communication link. It is defined as the probability that the received signal-to-noise ratio (SNR) falls below a predefined threshold, which corresponds to a target data rate $R$. The instantaneous SNR at the output of the FAS is $\gamma = P_S(\psi^*)^2/\sigma_n^2$. Thus, the outage probability is given by
\begin{align}\label{b1}
P_{\rm{out}}&=\Pr\left( \log_2(1+\gamma) \leq R \right) = \Pr\left(\psi^* \leq \Lambda_{\rm{th}}\right) \notag\\
&= F_{\psi^*}(\Lambda_{\rm{th}}),
\end{align}
where $F_{\psi^*}(\cdot)$ is the CDF of the selected channel gain $\psi^*$, and $\Lambda_{\rm{th}}=\sqrt{(2^R-1)\sigma_n^2/P_S}$. {Since $\gamma = P_S(\psi^*)^2/\sigma_n^2$, this threshold enforces the SNR condition $\gamma \ge 2^{R/W}-1$, meaning the cascaded FAS--RIS channel must provide a gain of at least $\Lambda_{\rm th}$ to sustain the target rate $R$ over bandwidth $W$.}

However, deriving a closed-form expression for $F_{\psi^*}(y)$ is intractable. This difficulty stems from two main challenges: (1) the complicated distribution of each channel gain $\psi_k$, which is a sum of products of Rayleigh RVs, and (2) the complex correlation structure among the channel gains $\{\psi_k\}_{k=1}^K$. To overcome these challenges and unlock a tractable analysis, we propose a two-pronged approximation strategy. First, we approximate the marginal distribution of each $\psi_k$ using a Gamma distribution. Second, we employ a block-correlation model to simplify the joint distribution of the channel vector $\mathbf{\Psi}$.

\subsection{Gamma Approximation for the Marginal Distribution}
The channel gain $\psi_k$ is a sum of $M$ i.i.d. RVs. According to the CLT, its distribution approaches a Gaussian distribution as $M \to \infty$. However, as motivated in Section I, practical vehicular networks often feature a small or moderate number of RIS elements, where the CLT approximation becomes inaccurate.

To develop a more accurate model, we approximate the distribution of $\psi_k$ with a Gamma distribution, which is well-suited for modeling sums of non-negative RVs, even for a small $M$. By applying the moment matching method, we match the first two moments of $\psi_k$ (given in \eqref{c1} and \eqref{c2}) with those of a Gamma distribution $\ga(a_k, b_k)$. {Moment matching preserves the exact first- and second-order statistics of the aggregated gain while keeping the Gamma form tractable; empirical tests confirm that higher-order discrepancies remain negligible across the simulated FAS configurations.} The shape parameter $a_k$ and scale parameter $b_k$ are found to be
\begin{align}\label{eq:gamma_params}
a_k=\frac{E_{\psi}^2}{V_{\psi}} = \frac{M\pi^2}{16(1-\pi^2/16)}, \quad b_k=\frac{V_{\psi}}{E_{\psi}} = \frac{4(1-\pi^2/16)\sqrt{\epsilon_1\epsilon_2}}{\pi}.
\end{align}
Consequently, the CDF of $\psi_k$ can be accurately approximated by
\begin{align}
F_{\psi_k}(x)\approx\frac{1}{\Gamma(a_k)}\gamma\left(a_k,\frac{x}{b_k}\right),
\end{align}
where $\gamma(\cdot,\cdot)$ is the lower incomplete gamma function, defined as $\gamma(a,x)=\int_{0}^{x}t^{a-1}e^{-t}dt$.
\begin{remark} 
The effectiveness of the Gamma approximation for a finite number of RIS elements $M$ can be understood by examining the shape parameter $a_k$. From \eqref{eq:gamma_params}, we see that $a_k$ is directly proportional to $M$. For a Gamma distribution to resemble a Gaussian distribution (where the CLT holds), its shape parameter must be large. When $M$ is small, as is common in practical vehicular deployments, $a_k$ is also small. In this regime, the Gamma distribution exhibits significant skewness and is fundamentally different from a symmetric Gaussian distribution. For instance, with $M=5$, $a_k \approx 5.7$, resulting in a PDF that is far from bell-shaped. This inherent structural mismatch is why the CLT-based approach leads to significant errors for small $M$. Conversely, our Gamma-based framework, by construction, accurately captures the distribution's shape even for small $M$, thus unlocking a realistic performance analysis and directly addressing the key contribution highlighted in Section I. {In addition, approximating $\psi_k$ with a Gaussian would allocate non-zero probability to negative channel gains and underestimate the outage tail, whereas the Gamma family preserves the support on $\mathbb{R}_+$, matches the first two cumulants via moment matching, and provides closed-form CDF/MGF expressions that we later leverage in the outage derivations. Compared with lognormal or Weibull alternatives, the Gamma model aligns with additive scattering physics (sums of Rayleigh products) and offers closed-form Laplace transforms, simplifying both outage and ergodic rate analysis.} As $M$ grows large, $a_k$ also becomes large, and the Gamma distribution itself converges to a Gaussian distribution, explaining why both methods yield similar results in the asymptotic regime.
\end{remark}

\subsection{Block-Correlation Model for Tractable Dependencies}
While the Gamma approximation accurately models the marginal distribution, the correlation among the $\{\psi_k\}$ still makes the analysis of their maximum, $\psi^*$, intractable. To address this, we simplify the correlation structure using the block-correlation model proposed in \cite{BC24}.

This model approximates the true, complex correlation matrix $\mathbf{\Omega}$ with a structured matrix $\hat{\mathbf{\Omega}}$ that is more amenable to analysis. Specifically, the set of $K$ correlated RVs $\mathbf{\Psi}$ is partitioned into $D$ groups. Within each group $d$, the $L_d$ RVs are equi-correlated with coefficient $\rb = \eta(\mu_d)$. Across any two different groups, the RVs are also equi-correlated, but with a different coefficient $\ra=\eta(0)$. This leads to a block-structured correlation matrix $\mathbf{\hat{\Omega}}$ for the approximated channel vector $\hat{\mathbf{\Psi}}=[\hat{\psi}_{1}, \cdots, \hat{\psi}_K ]^T$:
\begin{align}\label{d4}
\mathbf{\hat{\Omega}}\in\mathbb{R}^{K\times K}=\left(
         \begin{array}{cccc}
         \mathbf{B}_1 & \mathbf{C} & \cdots & \mathbf{C} \\
        \mathbf{C} & \mathbf{B}_2 & \cdots & \mathbf{C} \\
        \vdots &  & \ddots & \vdots \\
        \mathbf{C} & \mathbf{C} & \cdots & \mathbf{B}_D \\
         \end{array}
       \right).
\end{align}
Here, $\mathbf{B}_d \in \mathbb{R}^{L_d \times L_d}$ is a matrix with ones on the diagonal and $\rb$ elsewhere, and $\mathbf{C} \in \mathbb{R}^{L_d \times L_j}$ is a matrix with all entries equal to $\ra$. The parameters $L_d$ and $\mu_d$ are determined by minimizing the distance between the eigenvalues of the true and approximate covariance matrices, as detailed in \cite{BC24}.

A key benefit of this structure is that the RVs $\{\hat{\psi}_{l,d}\}$ within this model can be constructed from a set of common, independent Gamma RVs. Specifically, each $\hat{\psi}_{l,d}$ can be expressed as
\begin{align}
\hat{\psi}_{l,d}\approx r_{l,d}+w_d+t,
\end{align}
where $r_{l,d}\sim \ga((1-\rho_d)a_k, b_k)$, $w_d\sim \ga((\rho_d-\rho_0)a_k, b_k)$, and $t\sim\ga(\rho_0 a_k, b_k)$ are all mutually independent Gamma-distributed RVs. This decomposition is crucial for deriving the CDF of the maximum channel gain.

\subsection{Derivation of the Approximate Outage Probability}
With the Gamma approximation for the marginals and the block-correlation model for the dependencies, we can now derive a tractable expression for the outage probability. The CDF of the selected channel gain $\hat{\psi}^* = \max_{d,l} \hat{\psi}_{l,d}$ can be expressed by conditioning on the common RV $t$:
\begin{align}\label{d7}
F_{\hat{\psi}^*}(y) &= \mathbf{E}_t\bigg[\Pr\left(\max_{d,l} (r_{l,d}+w_d) \leq y-t \ | \ t\right)\bigg] \nonumber\\
&= \mathbf{E}_t\bigg[\prod_{d=1}^{D}F_{\hat{\psi}_d^*|t}(y)\bigg],
\end{align}
where $\hat{\psi}_d^*= \max_{1\leq l\leq L_d} \hat{\psi}_{l,d}$, and the product form in the second step is due to the independence among blocks after conditioning on $t$.

The conditional CDF $F_{\hat{\psi}_d^*|t}(y)$ can be found by further conditioning on $w_d$. For the case $\rho_d<1$, this leads to
\begin{align}
F_{\hat{\psi}_d^*|t}^2(y)&=\int_0^{y-t}\left[\frac{1}{\Gamma(\alpha_r)}\gamma\left(\alpha_r,\frac{y-t-x}{b_k}\right) \right]^{L_d} f_{w_d}(x) dx,
\end{align}
where $\alpha_r=(1-\rho_d)a_k$ and $f_{w_d}(x)$ is the PDF of the Gamma RV $w_d$. For the special case $\rho_d=1$, the expression simplifies to
$F_{\hat{\psi}_d^*|t}^1(y)=\frac{1}{\Gamma(\alpha_w)}\gamma(\alpha_w,\frac{y-t}{b_k})$, with $\alpha_w=(1-\rho_0)a_k$.

The integral in the $\rho_d<1$ case does not have a simple closed form. To obtain a final expression, we resort to the Gauss-Chebyshev quadrature method for numerical integration. This allows us to approximate the integral as a weighted sum. Applying this method to both the inner integral over $w_d$ and the outer expectation over $t$ yields the final closed-form approximate expression for the outage probability:
\begin{align}\label{d10}
P_{\rm out} = F_{\hat{\psi}^*}(\Lambda_{\rm th}) \approx \frac{\Lambda_{\rm th}\pi}{2(U_j+1)}\sum_{j=0}^{U_j}\sqrt{1-\chi_j^2}\prod_{d=1}^{D}F_{\hat{\psi}_d^*|t=\dj}(\Lambda_{\rm th}),
\end{align}
where $U_j$ is a parameter controlling the accuracy of the quadrature, and
\begin{align}
\chi_j=\cos\Big(\frac{(2j+1)\pi}{2U_j+2}\Big), \quad \dj=\frac{\Lambda_{\rm th}(\chi_j+1)}{2}.
\end{align}
The term $F_{\hat{\psi}_d^*|t=\dj}(\Lambda_{\rm th})$ is also computed using a similar Gauss-Chebyshev approximation. This final expression, while seemingly complex, is easily computable and provides the accurate outage performance analysis that was previously elusive for FAS-RIS systems with a finite number of elements.

\section{Numerical Results}
In this section, we report numerical and simulation results for the outage performance of the proposed FAS--RIS system. Unless otherwise stated, we set $\epsilon_1=\epsilon_2=1$ and choose the threshold as $\Lambda_{\rm th}=E_{\psi}$ per \eqref{c1}. The RIS applies optimal co-phasing as in \eqref{a2}. As a benchmark, we use Monte Carlo simulations under the Jakes' correlation model in \eqref{a4} with port selection in \eqref{a3} (denoted as ``Exact, Jakes''). Three analytical/approximate curves are compared: ``Gaussian-FBC'' refers to the CLT-based analysis with {fixed block-correlation (FBC)} where $\rho_d$ is treated as a constant (we set $\rho_d=0.97$); ``Gaussian-VBC'' refers to the CLT-based analysis with {variable block-correlation (VBC)} where $\rho_d$ for $d\in\{1,\dots,D\}$ is obtained via a heuristic procedure akin to \cite{BC24}{---specifically, the Gaussian-VBC baseline (i) approximates the cascaded channel as Gaussian via the CLT under rich scattering and (ii) adaptively selects correlation coefficients $\rho_d$ by matching eigenvalue spectra of the true and block-approximated covariance matrices}; ``Gamma (proposed)'' refers to the Gamma-based analysis developed in Section~III.

\begin{figure}[htbp]
  \centering
  \includegraphics[width=0.9\linewidth]{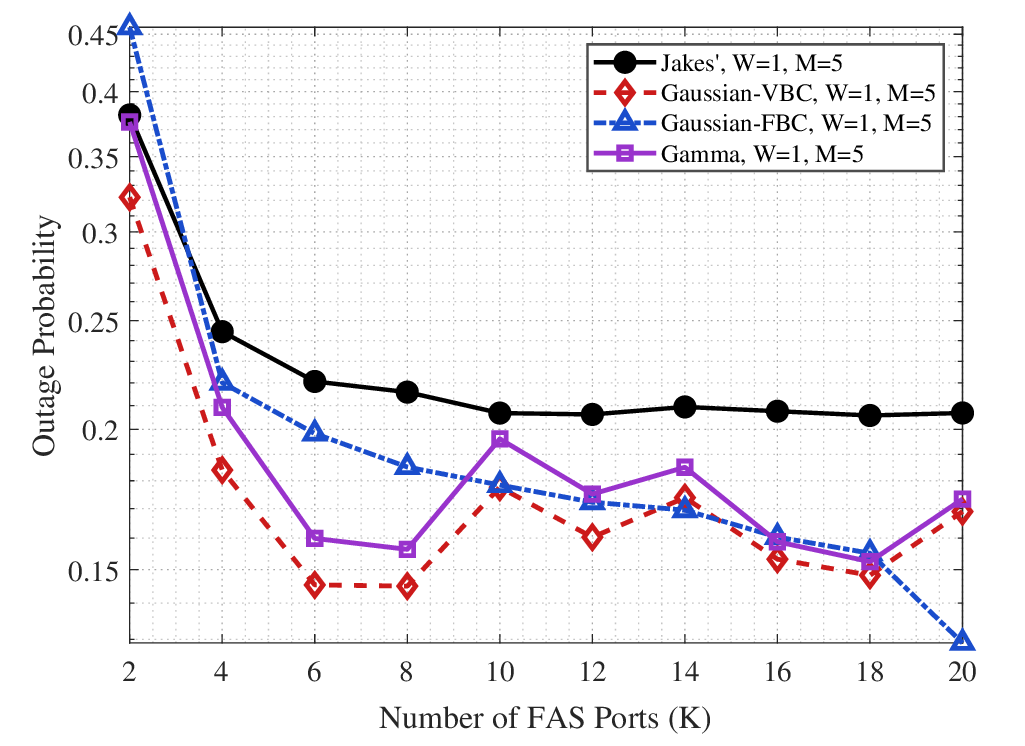}\\
  \caption{ {Outage probability versus the number of FAS ports $K$ (parameters: $W=1$, $M=5$).}}\label{fig2}
\end{figure}

Fig.~\ref{fig2} depicts $P_{\rm out}$ versus $K$ for $W=1$ and $M=5$. The proposed Gamma-based curve closely follows the ``Exact, Jakes'" benchmark and consistently outperforms both Gaussian baselines. The performance gain over CLT-based approaches is most pronounced for small $K$, highlighting the practical advantage of the proposed method for compact FAS devices in V2X scenarios.

\begin{figure}[htbp]
  \centering
  \includegraphics[width=0.9\linewidth]{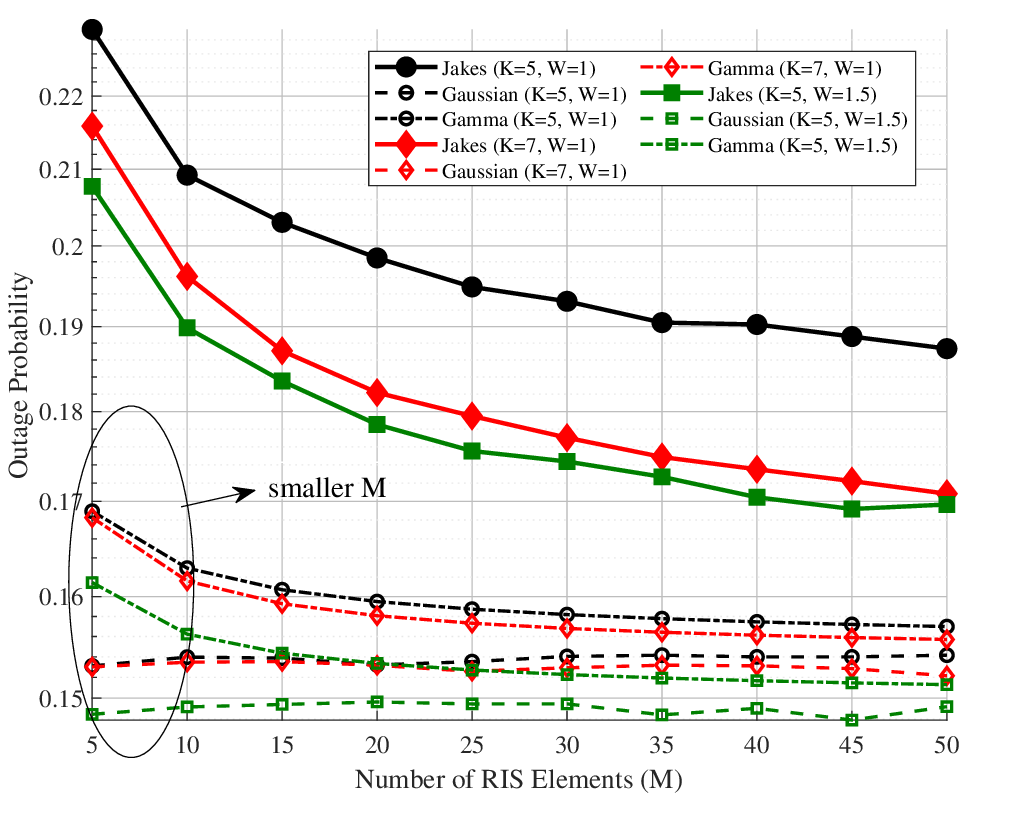}\\
  \caption{Outage probability versus the number of RIS elements $M$ (parameters: $W\in\{1,1.5\}$, $K\in\{5,7\}$).}\label{fig3}
\end{figure}

Fig.~\ref{fig3} plots $P_{\rm out}$ versus $M$ for $W\in\{1,1.5\}$ and $K\in\{5,7\}$. As $M$ increases, all approximations approach the benchmark, which is expected from the law of large numbers. For small and moderate $M$, the Gamma-based analysis remains much closer to the benchmark than the CLT-based curves. When $M$ exceeds roughly a few tens (e.g., $M\gtrsim 30$), the gap between Gamma- and Gaussian-based results narrows, aligning with the discussion in Remark~1.

\begin{figure}[htbp]
  \centering
  \includegraphics[width=0.9\linewidth]{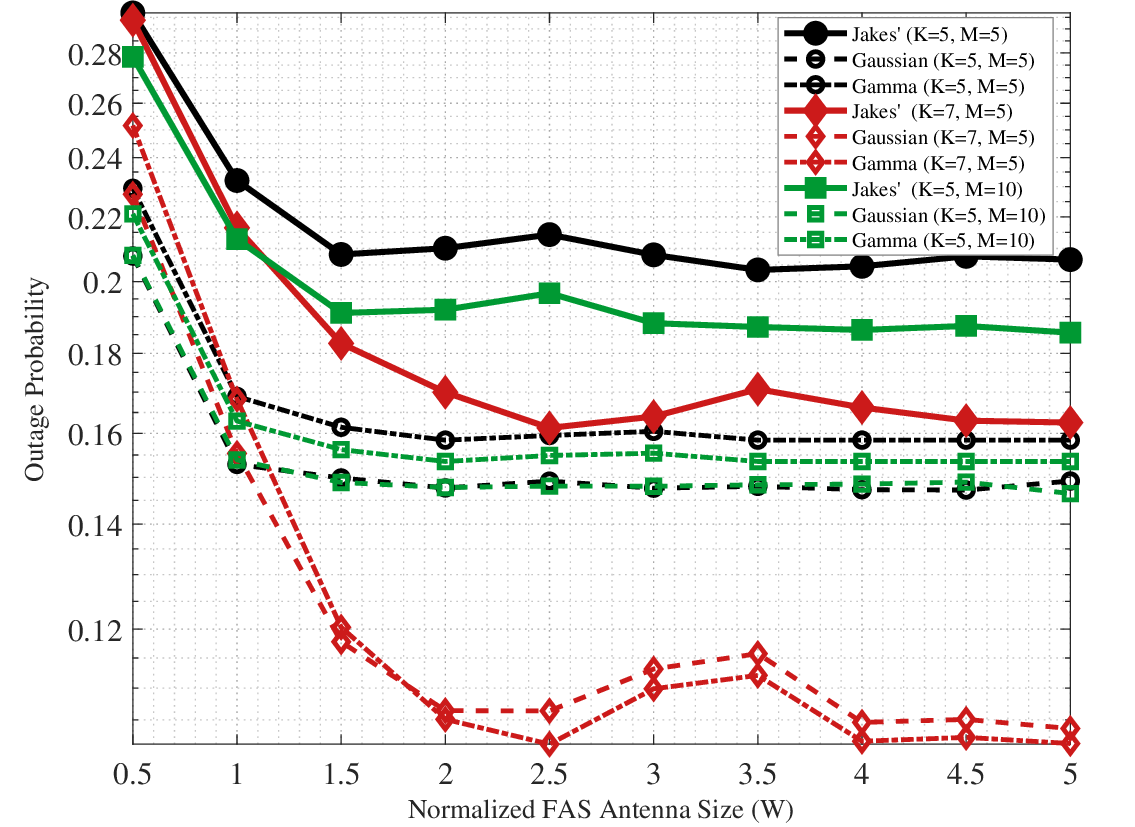}\\
  \caption{Outage probability versus normalized FAS aperture $W$ (parameters: $M\in\{5,10\}$, $K\in\{5,7\}$).}\label{fig4}
\end{figure}

Fig.~\ref{fig4} presents $P_{\rm out}$ versus the normalized aperture $W$ for $M\in\{5,10\}$ and $K\in\{5,7\}$. Increasing $W$ reduces inter-port correlation and improves selection diversity, thereby lowering $P_{\rm out}$. Across all $W$, the Gamma-based analysis tracks the benchmark more accurately than CLT-based methods, with the relative gain being larger when $K$ and $M$ are small.

\begin{remark}[V2X-oriented design]
For finite-size RIS and compact FAS in V2X: (i) use the proposed Gamma-based analysis for small/moderate $M$ and $K$ to avoid CLT-induced misestimation; (ii) increase $W$ when feasible to reduce inter-port correlation and improve selection diversity; (iii) under limited $M$, prioritize accurate phase control/placement over merely increasing $K$. For sizing to a target outage, map rate/SNR to $\Lambda_{\rm th}$ via \eqref{b1} and choose $K,M,W$ using the Gamma curves with margin. {Moreover, the derived outage expressions provide gradient cues for greedy or learning-assisted FAS port selection algorithms, enabling controllers to prioritize port subsets that jointly elevate the Gamma shape parameter and thereby tighten reliability guarantees.}
\end{remark}

\section{Conclusions}
We developed a Gamma-based outage analysis for FAS--RIS systems with finite $M$ and derived a tractable expression via block-correlation modeling and Gauss--Chebyshev quadrature. Numerical results show that the proposed analysis agrees closely with simulations and is markedly more accurate than CLT-based baselines, especially for small $K$ and $M$ (e.g., $M<30$), which is the practically relevant regime for V2X deployments. {Looking ahead, the framework can be extended to scenarios with line-of-sight components—including weak direct links and Rician fading—by incorporating the deterministic component via moment-matching techniques, yielding either a shifted-Gamma or a mixture model while preserving analytical tractability. Additionally, future work will incorporate imperfect CSI by propagating estimation error variances into the matched-Gamma parameters, allowing reliability margins to be quantified for practical pilot-limited V2X links.}



\begin{thebibliography}{1}




\bibitem{Wong-frontiers22}
K. K. Wong, {\em et al.}, ``Bruce Lee-inspired fluid antenna system: Six research topics and the potentials for 6G,'' {\em Frontiers Commun. Netw., section Wireless Commun.}, 3:853416, Mar. 2022.

\bibitem{MFAS23}K.-K. Wong, {\em et al.},  ``Fluid antenna system-part II: Research opportunities," \emph{IEEE Commun. Lett.}, vol. 27, no. 8, pp. 1924--1928, Aug. 2023.

\bibitem{TWu25}
T. Wu {\em et al.}, ``Fluid antenna systems enabling 6G: Principles, applications, and research directions," \emph{IEEE Wireless Commun. }, early access, 2025, doi: 10.1109/MWC.2025.3629597.

\bibitem{FAS20}K.-K. Wong, A. Shojaeifard, K.-F. Tong and Y. Zhang, ``Performance limits of fluid antenna systems," \emph{IEEE Commun. Letters}, vol. 24, no. 11, pp. 2469--2472, Nov. 2020.

\bibitem{Rodrigo14}D. Rodrigo, B. A. Cetiner, and L. Jofre, ``Frequency, radiation pattern and polarization reconfigurable antenna using a parasitic pixel layer," \emph{IEEE Trans. Antennas Propag.}, vol. 62, no. 6, pp. 3422--3427, Jun. 2014.

\bibitem{Huang21}Y. Huang, L. Xing, C. Song, S. Wang, and F. Elhouni, ``Liquid antennas: Past, present and future," \emph{IEEE Open J. Antennas Propag.}, vol. 2, pp. 473--487, 2021.

    \bibitem{Chai22}Z. Chai, K.-K. Wong, K.-F. Tong, Y. Chen, and Y. Zhang, ``Port selection for fluid antenna systems," \emph{IEEE Commun. Lett.}, vol. 26, no. 5, pp. 1180--1184, May 2022.

\bibitem{Waqar23}N. Waqar, K.-K. Wong, K.-F. Tong, A. Sharples, and Y. Zhang, ``Deep learning enabled slow fluid antenna multiple access," \emph{IEEE Commun. Lett.}, vol. 27, no. 3, pp. 861--865, March 2023.

\bibitem{Yang25}
S. Yang {\em et al.}, ``Toward intelligent antenna positioning: Leveraging DRL for FAS-aided ISAC systems," \emph{IEEE Internet Things J.}, vol. 12, no. 16, pp. 34615--34618, Aug. 2025.

\bibitem{Zhang25JSAC}
Z. Zhang, K.-K. Wong, J. Dang, Z. Zhang and C.-B. Chae, ``On fundamental limits for fluid antenna-assisted integrated sensing and communications for unsourced random access," \emph{IEEE J. Sel. Areas Commun.}, doi: 10.1109/JSAC.2025.3608113, 2025.

\bibitem{Zhang25WCL}
Z. Zhang, K.-K. Wong, J. Dang, Z. Zhang, C. Masouros and C.-B. Chae, ``On fundamental limits of slow-fluid antenna multiple access for unsourced random access," \emph{IEEE Wireless Commun. Lett.}, doi: 10.1109/LWC.2025.3594112, 2025.





\bibitem{FAS21}K.-K. Wong, A. Shojaeifard, K.-F. Tong, and Y. Zhang, ``Fluid antenna systems," \emph{IEEE Trans. Wireless Commun.}, vol. 20, no. 3, pp. 1950--1962, March 2021.

\bibitem{FAS22}K.-K. Wong, K. F. Tong, Y. Chen, and Y. Zhang, ``Closed-form expressions for spatial correlation parameters for performance analysis of fluid antenna systems," \emph{IET Electron. Lett.}, vol. 58, no. 11, pp. 454--457, May 2022.






\bibitem{FAMS}K.-K. Wong and K.-F. Tong, ``Fluid antenna multiple access," \emph{IEEE Trans. Wireless Commun.}, vol. 21, no. 7, pp. 4801-C4815, Jul. 2022.


\bibitem{FAMS23}K.-K. Wong, D. Morales-Jimenez, K.-F. Tong, and C.-B. Chae, ``Slow fluid antenna multiple access," \emph{ IEEE Trans. Commun.}, vol. 71, no. 5, pp. 2831--2846, May 2023.


\bibitem{Khammassi23}M. Khammassi, A. Kammoun and M. -S. Alouini, ``A new analytical approximation of the fluid antenna system channel,'' {\em IEEE Trans. Wireless Commun.}, vol. 22, no. 12, pp. 8843--8858, Dec. 2023.

 \bibitem{BC24}P. Ram\'{i}rez-Espinosa, D. Morales-Jimenez, and K. K. Wong, ``A new spatial block-correlation model for fluid antenna systems," \emph{IEEE Trans. Wireless Commun.}, vol. 23, no. 11, pp. 15829--15843, Nov. 2024.
 
 \bibitem{Lai25JSAC}
 X. Lai {\em et al.}, ``Revisiting spatial block-correlation model for fluid antenna systems: From constant to variable correlations," \emph{IEEE J. Sel. Areas Commun.}, doi: 10.1109/JSAC.2025.3617021, 2025.

\bibitem{WuTuo25}
T. Wu {\em et al.}, ``Scalable fluid antenna systems: A new paradigm for array signal processing," {\em arXiv preprint arXiv:2508.10831}, 2025.

\bibitem{Zheng25}
J. Zheng {\em et al.}, ``Unlocking FAS-RIS security analysis with block-correlation model," \emph{IEEE Wireless Commun. Lett.}, vol. 14, no. 7, pp. 2029--2033, Jul. 2025.








\bibitem{Yang20}L. Yang, Y. Yang, M. O. Hasna, and M. -S. Alouini, ``Coverage, probability of SNR gain, and DOR analysis of RIS-aided communication systems,"  \emph{IEEE Wireless Commun. Lett.}, vol. 9, no. 8, pp. 1268--1272, Aug. 2020.

\bibitem{Gan21}X. Gan, C. Zhong, Y. zhu and Z Zhong, ``User selection in reconfigurable intelligent surface assisted communication systems," \emph{IEEE Commun. Lett.}, vol. 25, no. 4, pp. 1353--1357, Appr. 2020.

\bibitem{ShiProc}
E. Shi, J. Zhang, H. Du, B. Ai, C. Yuen, D. Niyato and K. B. Letaief, ``RIS-aided cell-free massive MIMO systems for 6G: Fundamentals, system design, and applications," \emph{Proc. IEEE}, vol. 112, no. 4, pp. 331--364, Apr. 2024.

\bibitem{Shi25TWC}
E. Shi, J. Zhang, J. An, G. Zhang, Z. Liu, C. Yuen and B. Ai, ``Joint AP-UE association and precoding for SIM-aided cell-free massive MIMO systems," \emph{IEEE Trans. Wireless Commun.}, vol. 24, no. 6, pp. 5352--5367, Jun. 2025.

\bibitem{LaiX242} X. Lai, J. Yao, K. Zhi, T. Wu, D. Morales-Jimenez, and  K. K. Wong,  ``FAS-RIS: A block-correlation model analysis," {\em IEEE Trans. Veh. Technol.}, vol. 74, no. 2, pp. 3412--3417, Feb. 2025.

\bibitem{YaoJ251} J. Yao, {\em et al.}, ``FAS-RIS communication: Model, analysis, and optimization," {\em IEEE Trans. Veh. Technol.}, vol. 74, no. 6, pp. 9938--9943, Jun. 2025.

\bibitem{Wu25RIS}
T. Wu {\em et al.}, ``Exploit high-dimensional RIS information to localization: What is the impact of faulty element?" \emph{IEEE J. Sel. Areas Commun.}, vol. 42, no. 10, pp. 2803--2819, Oct. 2024.



\bibitem{NumericalAnalysis} E. Suli and D. F. Mayers, \emph{An Introduction to Numerical Ananlysis}. Cambridge, U.K.: Cambridge Univ. Press, 2003.

\end{thebibliography}
\end{document}